# Pellet refuelling of particle loss due to ELM mitigation with RMPs in the ASDEX Upgrade tokamak at low collisionality

M Valovič[1], P T Lang[2], A Kirk[1], W Suttrop[2], M Cavedon[2], G Cseh[3],
L R Fischer[2], L Garzotti[1], L Guimarais[2], G Kocsis[3], B Plőckl[2], R Scannell[1],
T Szepesi[3], A Thornton[1], A Mlynek[2], G Tardini[2], E Viezzer[2], E Wolfrum[2], the ASDEX Upgrade team[2] and the EUROfusion MST1 team[4]

[1]CCFE, Culham Science Centre, Abingdon, OX14 3DB, UK
[2]Max-Planck-Institut für Plasmaphysik, Boltzmannstrasse 2, D-85748 Garching, Germany
[3]Wigner Research Centre for Physics, HAS, Budapest, Hungary
[4]see http://www.euro-fusionscipub.org/mst1

E-mail: martin.valovic@ccfe.ac.uk

**Abstract.** The complete refuelling of the plasma density loss (pump-out) caused by mitigation of Edge Localised Modes (ELMs) is demonstrated on the ASDEX Upgrade tokamak. The plasma is refuelled by injection of frozen deuterium pellets and ELMs are mitigated by external resonant magnetic perturbations (RMPs). In this experiment relevant dimensionless parameters, such as relative pellet size, relative RMP amplitude and pedestal collisionality are kept at the ITER like values. Refuelling of density pump out of the size of $\Delta n / n \sim 30\%$ requires a factor of two increase of nominal fuelling rate. Energy confinement and pedestal temperatures are not restored to pre-RMP values by pellet refuelling.

## 1. Introduction

The operation of tokamak fusion reactors is based on plasmas in the high confinement regime, the H-mode. However the H-mode is typically accompanied by Edge Localised Modes (ELMs) which are not compatible with the long term divertor operation and therefore ELM control has to be considered [1]. One of the ELM control techniques is the application of Resonant Magnetic Perturbations (RMPs) produced by an external set of magnetic coils [2, 3, 4, 5] and such a system will be installed on ITER.

The application of RMPs in tokamaks has, however, unwanted side effects. It is often observed that when ELMs are mitigated the plasma density is significantly reduced which is dubbed as a "density pump-out" effect. This happens if RMPs increase inter-ELM transport or the increase of ELM frequency $f_{ELM}$ by RMPs is not matched by sufficient reduction of particle loss per ELM $\delta N_{ELM}$ and thus the related particle loss $\Phi_{ELM} = f_{ELM} \delta N_{ELM}$ increases.

In ITER, the plasma density should be carefully adjusted during all phases of plasma evolution as it is the density through which the fusion power is controlled [6]. During the H-mode phase the density control by gas fuelling is likely to be inadequate [7] and fuelling by frozen deuterium-tritium pellets launched from the high field side of the plasma is part of the ITER design. As a consequence, ELM control by RMPs and pellet fuelling should be tested simultaneously as a part of integrated scenario development.

Simultaneous pellet fuelling and ELM control by RMPs have been attempted on a number of machines. In DIII-D, the plasma with ELMs suppressed by RMPs has been refuelled by pellets but at a cost of return to ELMy H-mode [8]. On JET, pellets have been used to refuel plasmas with RMPs though at relatively low plasma currents and using low field side pellets [5]. In ASDEX Upgrade, compatibility of pellet fuelling with ELM suppression by RMPs has been demonstrated although at high plasma collisionality and deeper pellet deposition [9]. On MAST, plasmas with RMPs have been refuelled by high field side pellets with moderate effect on ELM mitigation but gas fuelling was significant and $\delta N_{ELM}$ ~3% of the plasma particle content, i.e. ~6 times larger than the ITER target [10, 11].

The present experiment was performed on ASDEX Upgrade and was designed specifically to demonstrate simultaneous density and ELM control under conditions envisaged in ITER during the density ramp and flat top H-mode phases [6]. The plasma collisionality, RMP amplitude and pellet relative size are set close to the values expected in ITER (for numerical values see the sections below). We decided that the parameter to restore after application of ELM control is the plasma density as this will be the control situation during the density ramp and flat top phases in ITER. Another possibility would be to restore the plasma temperature but this would not mimic the ITER situation where during the density ramp up phase full auxiliary power is necessary and during the flat top phase fusion power will dominate.

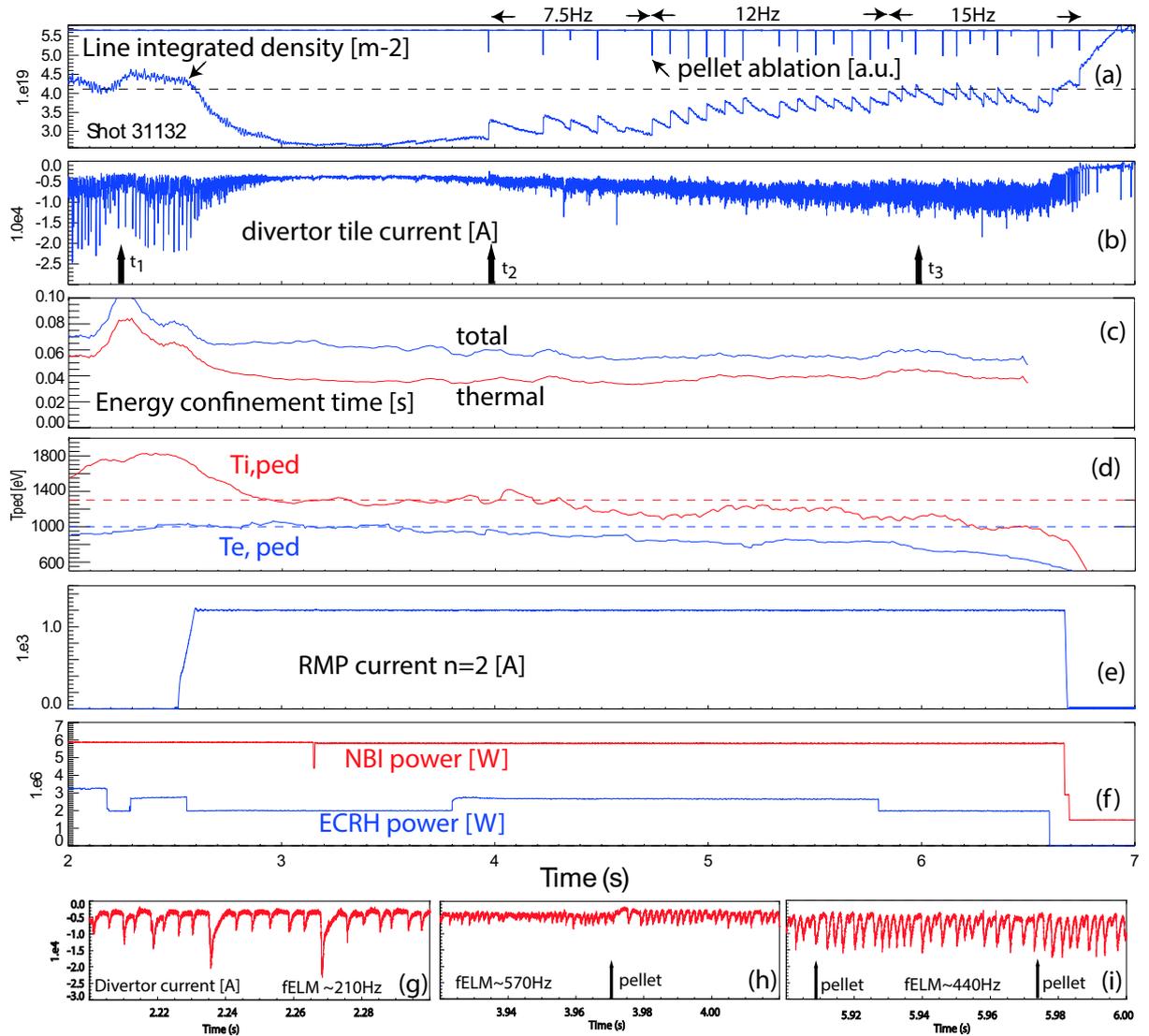

**Figure 1.** Temporal evolution of plasma parameters with ELM mitigation by RMPs and pellet fuelling. (a) line integrated density (bottom trace), pellet ablation radiation monitor (top trace, inverted), (b) divertor strike point current, (c) energy confinement times, (d) time averaged electron and ion pedestal temperatures are calculated as: $T_{e,ped} = \langle T_e(\rho \approx 0.94) \rangle_{\Delta t=0.3s}$, $T_{i,ped} = \langle T_i(\rho = 0.93) \rangle_{\Delta t=0.1s}$, (e) RMP current, (f) NBI and ECRH power. Panels (g)-(i): temporal zoom into divertor strike point current at times indicated by arrows on panel (b).

## 2. Experimental setup

The experiment was performed in the ASDEX Upgrade tokamak. The plasma has a single null divertor, with radius of the geometric axis $R_{geo} = 1.62m$, minor radius $a = 0.482m$, plasma current $I_p = 0.82MA$, toroidal field $B_T(R = 1.65m) = 1.83T$ and safety factor $q_{95} = 3.8$. Fresh boronisation is applied to obtain low density and consequently low plasma collisionality. To improve reliability of the discharge small gas fuelling is applied with constant rate of $\Phi_{gas} = 0.5 \times 10^{21} at/s$. Traces of the plasma parameters are shown on figure 1. The plasma is heated mainly by neutral beams. In addition smaller amount of electron cyclotron heating (ECRH) is added with 3rd harmonic resonance absorption on axis. There is some residual power absorbed at 2$^{nd}$ harmonics layer located at top of the pedestal $\rho_{pol} = 0.85 - 0.90$ on the high field side ($\rho_{pol} = \sqrt{\psi_N}$ where $\psi_N$ is the normalised poloidal magnetic flux). Simulation by TORBEAM code [12] shows that this residual 2$^{nd}$ harmonic power increases during the shot due to the gradual decrease of electron temperature and at ~6.0s it reaches 0.45MW. This is about 5% of total heating power and thus the residual 2$^{nd}$ harmonic power absorbed at the pedestal top should not significantly affect the discharge scenario.

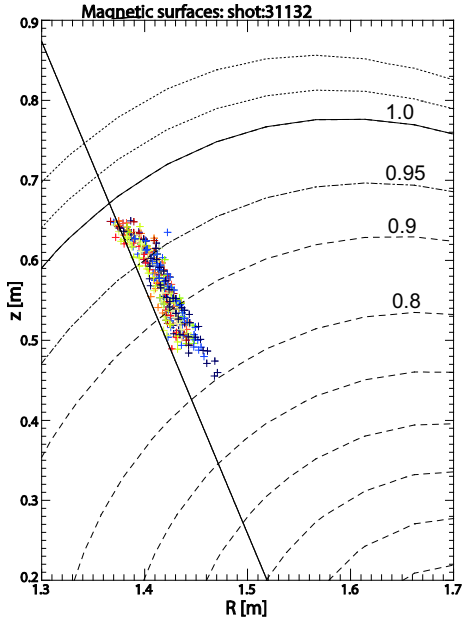 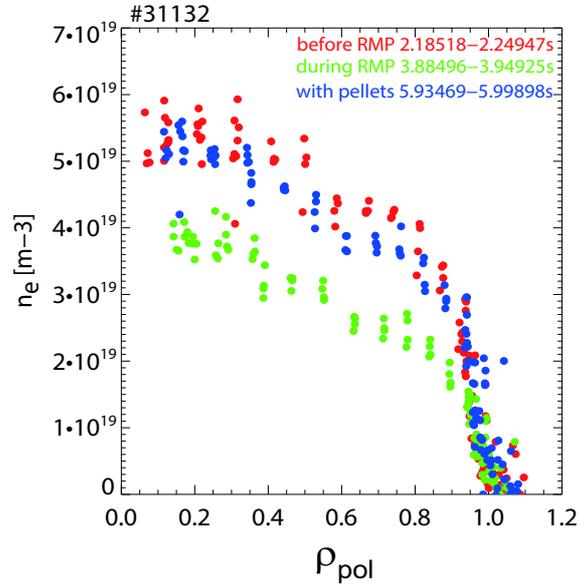

**Figure 2.** Ablation traces for all pellets in figure 1. Each point represents pellet position at one frame of visible light camera with 75000 frames/s and $6.5 \mu s$ exposure time. Labels correspond to the surfaces $\rho_{pol} = const$.

**Figure 3.** Density profiles before RMPs (red), during RMPs (green) and after pellet re-fuelling (blue). The times around which the profiles are taken are indicated by arrows in figure 1b. For each time point the figure shows about four consecutive Thomson scattering profiles within the specified time intervals, without averaging.

ELM mitigation is provided by the RMP field created by an array of 8 upper and 8 lower coils located at the low field side of the plasma. Coils are connected to create a perturbation with $n = 2$ toroidal mode number and with 90 degree spatial phase-shift between lower and upper array for magnetic field alignment. For more details of this scenario see the reference [13].

Switching on the ELM mitigation coils causes the density drop (see figure 1a). In addition RMPs also cause the drop of ion pedestal temperature while the electron pedestal temperature remains relatively unaffected. These are typical features of ELM mitigation by RMPs at low collisionality and are discussed in detail in reference [13]. Note that the reference phase just before RMPs is not fully stationary because of unintended drop of ECRH power between 2.2s and 2.4s (see figure 1f). As a result there is a sudden increase of density followed by the change in energy content and consequent transient in energy confinement time.

After the density reached a new quasi-stationary phase with RMPs the plasma is refuelled by pellets. The geometry of the pellet injection is shown in figure 2. Deuterium pellets have a nominal size of $1.4 \times 1.4 \times 1.5 mm$. Allowing 30% loss in the flight line the pellet particle content is $N_{pel} = 1.2 \times 10^{20}$ atoms. This gives within a factor of 2 the same pellet-to-plasma particle ratio as expected in ITER, i.e. 7% [7]. Pellets are launched from the high field side with a velocity of 560 m/s. Analysed fast camera images in visible light (figure 2) show that pellets are evaporated in the outer 20% of the minor radius ($\rho_{pol} > 0.8$). Due to the geometry of the pellet launch some redistribution of pellet material by $\nabla B$ - drift is expected but nevertheless pellet deposition partially mimics ITER-like shallow pellets. The pellet frequency is increased in 3 steps as seen in figure 1a. At the initial pellet rate $f_{pel} = 7.5 Hz$ the density changes only marginally. By doubling the pellet rate to $f_{pel} = 15 Hz$ the density increases and it is restored to pre-RMP values. Figure 3 shows the density profiles at 3 time points: $t_1$ - just before RMP application, $t_2$- after RMPs just before pellet fuelling and $t_3$- after pellet refuelling (see arrow markers in figure 1b). It is seen that after pellet refuelling the density profile is within ~10% the same as before application of RMP.

This shot is a result of several calibration shots in which the pellet size, frequency, velocity and timing were varied to obtain the best match in refuelling.

The critical question about refuelling of the density pump out is the price to pay in terms of efficiency of ELM mitigation, efficiency of pellet fuelling and possible reduction in confinement. These three aspects are now discussed separately below.

## 3. Efficiency of ELM mitigation

ELMs are detected by current on the tile at the divertor strike point. This signal is shown on figure 1b and in details around time points $t_1$, $t_2$, $t_3$ on panels (g), (h) and (i). Before the application of RMPs the ELM frequency is $f_{ELM}(t_1) = 210 Hz$ and it is a mixture of large (type-I) ELMs and smaller ELMs. Whether the smaller ELMs are type-IV (often referred as the low collisionality branch of type III ELMs [14]) remains to be identified. When RMPs are applied the ELM frequency increases to $f_{ELM}(t_2) = 570 Hz$. In this phase only small ELMs are present. The application of pellet refuelling leads to a decrease of ELM frequency to $f_{ELM}(t_3) = 440 Hz$. Despite this decrease, the ELM frequency after refuelling is still 2 times higher compared to the phase before RMP (time point $t_1$).

The relative energy loss per ELM before RMPs for large type-I ELMs is $\delta W_{ELM}/W(t_1) \approx 5\%$, as determined from equilibrium reconstruction ($W$ is the stored plasma energy). After the application of RMPs the ELM loss is so small that it is not reliably measured by equilibrium reconstruction and we use the inverse of normalised ELM frequency $\delta W_{ELM}/W \sim \left(f_{ELM}\tau_{E,th}\right)^{-1}$ as a proxy, where $\tau_{E,th}$ is the thermal energy confinement time calculated by TRANSP [19] (see figure 1c). Application of RMPs reduces the inverse of normalised ELM frequency from $\left(f_{ELM}\tau_{E,th}\right)^{-1}(t_1) \approx 7.2\%$ to $\left(f_{ELM}\tau_{E,th}\right)^{-1}(t_2) \approx 4.5\%$ (here $\tau_{E,th}(t_1)$ is an average over $\pm 0.25 s$). After refuelling by pellets inverse of normalised ELM frequency $\left(f_{ELM}\tau_{E,th}\right)^{-1}$ increases to $\left(f_{ELM}\tau_{E,th}\right)^{-1}(t_3) \approx 5.0\%$. Therefore refuelling of density pump out preserves ELM mitigation but at compromised level, with $\left(f_{ELM}\tau_{E,th}\right)^{-1}$ about 5 times larger than the ITER target $\left(f_{ELM}\tau_{E,th}\right)^{-1} \sim 1\%$ [7]. The infrared camera data show that the peak power density at divertor due to ELMs, $q_{peak}$, displays the similar trend: Application of RMPs reduces $q_{peak}$ approximately twofold. The subsequent pellet refuelling does not change the power density and within the data scatter $q_{peak}(t_2) \sim q_{peak}(t_3) \sim 4 MW/m^2$.

Density pump out raises the question whether the ELM mitigation is not simply the result of change of density and RMPs act only as a density control tool. Broadly speaking our data show that the ELM frequency is not a simple function of pedestal density as this is almost the same before RMPs and after refuelling by pellets, $n_{e,ped}[t_1,t_3]=[2.6,2.9]\times 10^{19}\,\text{m}^{-3}$, while the ELM frequency differ by a factor of two $f_{ELM}[t_1,t_3]=[210,440]\,\text{Hz}$. In addition if $f_{ELM}$ would be a simple function of density one would expect that the modulation by pellets will cause the synchronous modulation of ELM frequency and this is not observed as seen in figures 1h and 1i. The data also do not support simple dependence of $f_{ELM}$ with ion collisionality: Between $t_1$ and $t_3$ the ion collisionality increases as $\nu_{*i,ped}[t_1,t_3]=[0.089,0.21]$ suggesting $f_{ELM}$ is increasing with increasing $\nu_{*i,ped}$, however this trend is not supported by time point $t_2$ were $\nu_{*i,ped}[t_2]=0.10$, $f_{ELM}[t_2]=570\,Hz$, suggesting $f_{ELM}$ decreases with $\nu_{*i,ped}$ between $t_2$ and $t_3$. Here, $\nu_{*i,ped}=4.9\times 10^{-19}qR_{geo}n_{i,ped}[\text{m}^{-3}]Z^4\ln\Lambda_{ii}/\left(T_{i,ped}^2[\text{eV}]\varepsilon^{3/2}\right)$ [15], $\ln\Lambda_{ii}\approx 17$, $\varepsilon=a/R_{geo}$, $n_{e,ped}[t_2]=1.69\times 10^{19}\,\text{m}^{-3}$, $T_{i,ped}[t_1,t_2,t_3]=[1735,1264,1085]\,\text{eV}$, safety factor $q=3.3$, ion charge $Z=1.5$, $n_{i,ped}=n_{e,ped}/Z$. The fact that RMPs affect the ion and not the electron pedestal temperatures might indicate that it is the ion collisionality that might be relevant. Nevertheless the values of electron-ion collisionalities are similar $\nu_{*ei,ped}[t_1,t_2,t_3]=[0.18,0.11,0.25]$.

Finally note that even in a phase with full pellet refuelling ($t_3$) the ion collisionality is similar to that expected in ITER $\nu_{*i,ped}=0.19$ ($T_{i,ped}=4\,keV$, $n_{e,ped}=8\times 10^{19}\,m^{-3}$). This means that pellet fuelling in ASDEX Upgrade is compatible with the low collisionality regime relevant to ITER operation.

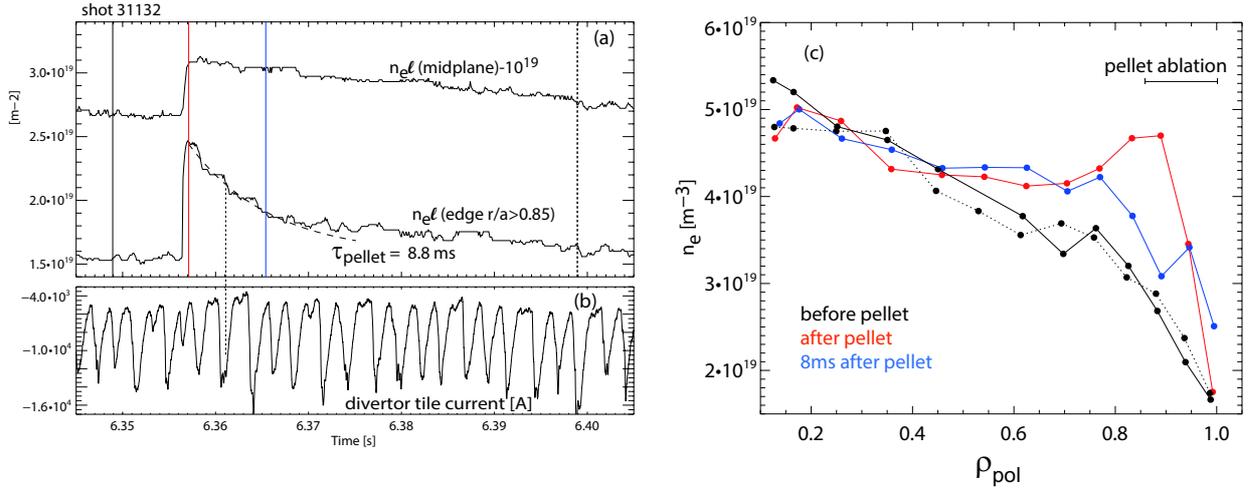

**Figure 4.** Detail of pellet deposition cycle. (a) core and edge interferometer signals, (b) divertor current signal, (c) density profiles from Thomson scattering at times indicated by vertical lines in panel (a) in corresponding colours and line style.

## 4. Fuelling rate

The density pump out indicates that RMPs increase particle loss. The size of this additional loss channel can be estimated from the time derivative of the electron particle content at the time of RMP initiation: $\Phi_{RMP}=dN_e/dt|_{t=2.63s}\approx d(n_e\ell)/dt|_{t=2.63s}V/2a\approx 1.0\times 10^{21}\,\text{at/s}$. Here $(n_e\ell)$ is the horizontal line integrated density, $V=12.1\,m^3$ is the plasma volume and $a$ is the horizontal plasma radius. After the transient phase lasting about 0.3s the plasma density equilibrates to about 60% of its pre RMP value. This decrease is approximately proportional over the whole plasma cross section from the pedestal to the core as seen in figure 3.

The above estimate for $\Phi_{RMP}$ is comparable to nominal fuelling rates: The pellet frequency at which the initial density is completely restored is $f_{pel}=15Hz$ giving the fuelling rate of $\Phi_{pel}=N_{pel}f_{pel}=1.9\times10^{21}$ at/s. The nominal gas valve flow rate is $\Phi_{gas}=0.5\times10^{21}$ at/s and the fuelling rate of neutral beams is $\Phi_{NBI}=0.88\times10^{21}$ at/s, both approximately constant in time. Combining these values the ratio of nominal fuelling rates before and after refuelling is: $\Phi(t_3)/\Phi(t_2)=1+\Phi_{pel}/(\Phi_{gas}+\Phi_{NBI})=2.4$. In other words the cost for refuelling of RMP pump out, of the size of 30%, is the increase of nominal fuelling rate by a factor of 2.4.

The fact that nominal fuelling ratio $\Phi(t_3)/\Phi(t_2)$ is larger than the ratio of electron particle contents after and before refuelling, $N_e(t_3)/N_e(t_2)$, is the combination of two effects. Firstly, the nominal gas valve flow rate $\Phi_{gas}$ is not equal nor proportional to the gas fuelling rate as the neutral pressure around the plasma is linked to recycling and can vary in time. (In our case the neutral gas density in the divertor roughly doubles, $(3.5\rightarrow6.2)\times10^{19}$ at/m$^3$, during the pellet refuelling phase $4\rightarrow6s$.) The second effect is that the global particle confinement time $\tau_p=N_e/\Phi$ for phase with pellet fuelling is different from the phase with gas and beam fuelling. This is because these three fuelling methods deposit particles at different plasma radii and local particle transport is different in corresponding parts of the plasma. In addition, for pellets the link between the global particle confinement time and the local particle transport is more complex because of the transient character of post-pellet particle losses. For a single pellet, at the fully refuelled phase, the whole pellet cycle is shown in figure 4. It is seen from the signals of line integrated densities that the post-pellet losses involve two phases: fast loss lasting ~10ms followed by slow density decay up to the next pellet. These two timescales are now discussed separately below.

*4.1 Fast time scale particle loss*
Figure 4a shows that the line integrated density covering the edge zone with $\rho_{pol}>0.85$ decays with a characteristic time constant of $\tau_{pellet}=8.8ms$. Thomson scattering density profiles taken at times which are bracketing this phase (red and blue symbols in figure 4a and 4c) show that 6.8% of total plasma particles is lost during this time interval (the corresponding change in the core line integrated density is 2.8%). The time constant $\tau_{pellet}$ is referred to as the pellet retention time. When normalised to the global energy confinement time $\tau_E$ (see figure 1c) the ratio is $\tau_{pellet}/\tau_E=0.19$. This value is similar to that measured in MAST with RMP ELM mitigation and shallow pellet deposition, $\tau_{pellet}/\tau_{E,}=0.17$ [11]. The nominal pellet diameter in MAST was 1.3mm, similar to that in ASDEX Upgrade.

From figure 4a it is also seen that this density decay is not continuous but occurs in steps which are well correlated with ELMs. There are about 3-4 ELMs responsible for the whole density loss during the $\tau_{pellet}$ timescale. Comparing the density profiles immediately after the pellet evaporation and 8ms after the pellet one can see that during this time interval a large fraction of the pellet material in the zone $\rho_{pol}>0.77$ is lost. It is also obvious that the character of the particle loss is not diffusive as the loss occurs also in the zone with initially inverted density gradient ($dn_e/d\rho_{pol}>0$). This observation is similar to that in MAST [11] where it was interpreted as being due to the existence of sizeable eddy structure during the ELM.

Comparing ELMs before and after the pellet one can see that ELMs are not significantly affected by pellets, at least as monitored by the divertor tile current (see figure 4 and figure 1(i)). This situation would be favourable in ITER where one of the concerns is that the pellet can trigger one or a burst of several ELMs that will result in prompt particle loss. Nevertheless this weak pellet-ELM interaction on ASDEX Upgrade is quite surprising given the large change in density gradient inside the separatrix due to the pellet, but could indicate that pedestal stability and strength of edge transport barrier is controlled mainly by pressure, and simultaneously the pellet deposition is close to adiabatic. In comparison in MAST RMP experiment, the post-pellet ELMs are about 1.5 times larger compared to pre-pellet ELMs [11].

Between ELMs the edge interferometer signal is approximately constant (see figure 4, at ~6.36 s). This is in contrast with the phase before application of RMPs (around time point $t_1$ when the density typically increases between large ELMs. This indicates that even immediately after the pellet, when the density gradient inside the separatrix is large, the status of the edge transport barrier is different compared to the pre RMP phase. Such a conclusion is supported by a measurement of the radial electric field (though time averaged) showing clear difference between pre RMP and pellet refuelled phases (see next section and figure 5). The above is also in line with our previous conclusion that RMPs are directly responsible for ELM control rather than the density itself.

The density profile immediately after the pellet is hollow (figure 4) raising a question of the intensity of inward particle transport. In the zone $\rho_{pol} = 0.45 - 0.65$, 8ms after the pellet, the density is slightly higher than the density immediately after the pellet perhaps hinting the existence of inward particle flux. However the analysis of this important process would require temporarily resolved profile data (or box car analysis) which is outside the scope of this paper. Post pellet plasmas with hollow density profiles were analysed in MAST in references [16, 17]. It was found that an inverted density gradient can unfavourably suppress micro-instabilities reducing the core fuelling rate, however, this depends on the actual values of the temperature gradients after the pellet. For completeness we note that the fast redistribution of density can be caused also by sawteeth which are present during the pellet refuelled phase. In our case, however, sawteeth are not correlated with pellets and thus not responsible for fast redistribution of plasma particles after the pellets.

*4.2 Slow time scale particle loss*
About 10 ms after the pellet the line integrated densities decay with the time constant much longer than the edge pellet retention time (figure 4). During this phase the density decreases in the outer part of the plasma $\rho_{pol} > 0.45$ while in the core remains constant so that the profiles become gradually more peaked. Comparison of profiles at the beginning and the end of the pellet cycle show good agreement confirming that the plasma is in a quasi-stationary phase. The fact that a quasi-stationary density is sustained with the pellet frequency much smaller than the inverse of the edge pellet retention time $1/\tau_{pellet}$ is a consequence of deeper pellet deposition. A fast camera data show that pellet ablation and ionisation occurs up to the normalised radius of $\rho_{pol} > 0.86$. However the pellet particles are deposited much deeper up to $\rho_{pol} \sim 0.45$ as seen in figure 4c. This difference can be atributed to $\nabla B$ drift of pellet material due to the high field side injection geometry. Such low pellet frequency would be favourable for ITER fuelling but the extrapolation from our data is not straightforward as the predicted pellet deposition is shallower in ITER meaning a shorter decay phase. Note that the separation between fast and slow decay phases is not sharp as the density profile evolves from hollow to peaked.

During the whole pellet cycle the density gradient scale length just inside the separatrix is comparable to the ion banana full width $\Delta r_b / a = 0.16$. This suggests that finite ion Larmor radius effects could be an important part of the mechanism of particle loss in these low collisionality pellet fuelled plasmas. The significance of this effect in ITER is however not obvious because the normalised ion Larmor radius is 6 times smaller in ITER compared to the present ASDEX Upgrade plasma.

Finally note that the evolution of density inside the separatrix over the whole pellet cycle illustrates the challenge of simultaneous pellet fuelling and ELM control by RMPs. On the one hand the fuelling pellet is increasing the density gradient in $\rho_{pol} > 0.9$ while after the pellet this density gradient is reduced. This gradual reduction of density gradient between pellets results from the fact that the density inside the separatrix is not in equilibrium and particle loss due to RMPs is larger than NBI and gas fuelling. This is different from the situations with gas fuelled plasmas where statements about the effect of RMPs on edge density gradient can be made, e.g. small decrease [18] or no effect [13]. In ITER effect of gas fuelling will be even smaller and therefore the evolution of density profiles between pellets will be mainly controlled by RMPs. From the control point of view we will be left with a task how to balance pellets and RMPs so that the density and ELMs are simultaneously acceptable over the whole pellet cycle.

## 5. Energy confinement

Application of RMPs is reducing the energy confinement time by ~ 30% (between $t_1$ and $t_2$ in figure 1c). This is the case for both the total value $\tau_{E,tot}$ (including fast ions) and thermal value $\tau_{E,th}$, where both quantities are calculated by TRANSP code [19]. During the refuelling phase by pellets (between $t_2$ and $t_3$) the energy confinement increases only by ~13%, clearly not restoring to the pre RMP value. In addition this increase is even less than predicted by IPB98(y,2) scaling, $\tau_{E,th} \propto n_e^{0.41}$ [20], according to which $\tau_{E,th}$ should increase by 16-20%. Finite beam shine through does not affect these conclusions as it is included in calculation of $\tau_{E,th}$ by TRANSP, and its variation is small (shine through reaches its maximum of $0.85\text{MW}$ at $3.1\text{s}$ and monotonically decreases to $0.3\text{MW}$ at $6.0\text{s}$).

The aforementioned global confinement broadly correlates with the behaviour at the pedestal. The application of RMPs reduces the pedestal ion temperature by 37% (figure 1d). Simultaneously the radial electric field at the pedestal becomes less negative by ~50% (figure 5). In contrast the electron pedestal temperature is not affected by the application of RMPs (see figure 1d). During the pellet refuelling phase, both electron and ion pedestal temperatures are modestly reduced, by 16% and 19% for ions and electrons resp. This change is somewhat lower than the increase of pedestal density during refuelling. The modest change in temperatures during refuelling is echoed by the fact that the radial electric field is unchanged within the error bars (figure 5).

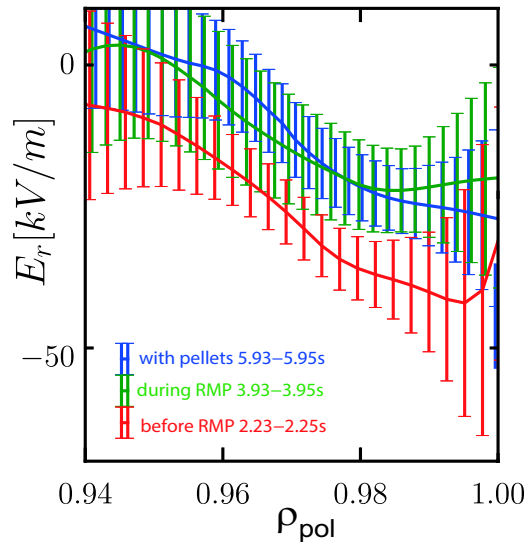

**Figure 5.** Radial electric field $E_r$ measured by boron 5 spectral line for different times of shot #31132: red before RMPs, green during RMPs and blue after pellet refuelling.

The correlation of density pump out and drop of ion temperature suggests that the particle and heat fluxes are connected. In analogy with the link between the plasma particle and heat transport in the plasma core [21] one can calculate the dimensionless ratio of the particle, $\Gamma_e$, and heat flux, $Q$, such as $\Gamma_e T/Q$, with $T$ being the plasma temperature. Applying this parameter to the pedestal top at the time of fully refuelled plasmas (time interval $t_3$ on figure 1) we find that the value linked to ion pedestal temperature is $\Gamma_e T_{i,ped}/Q \sim \Phi(t_3)T_{i,ped}(t_3)/P_{aux} = 0.05$, where $P_{aux}$ is the heating power. The numerical value of this parameter is controlled by the physics of transport mechanism. For example in case of convective ELMs, $\Delta N_{e,ELM}/N_{ped} = \Delta W_{ELM}/W_{ped}$, this ratio is $\Gamma_e T_e/Q \sim 0.33\alpha \sim 0.066$, where $\alpha \sim 0.2$ is the fraction of heat flux transported by ELMs [7]. Therefore this parameter provides a useful constraint for the plasma pedestal temperature, in particular it shows that, for a given power and character of pedestal transport, the pedestal temperature is inversely proportional to the fuelling flux. This illustrates the challenge of simultaneous ELM and density control when ELM mitigation at constant density results in increased fuelling which in turn leads to the reduction of pedestal temperature.

## 6. Conclusions

Complete refuelling of density pump out due to ELM mitigation by RMPs is demonstrated by pellets under conditions of ITER-like relative RMPs amplitude, pellet size and plasma collisionality. It is shown that:
- ELM mitigation is preserved by pellet refuelling. ELM frequency with pellets and RMPs is higher by a factor of 2 compared to non-RMP non-pellet reference (for normalised ELM frequency $f_{ELM}\tau_{E,th}$ this factor is 1.5)
- ELM frequency is not a simple function of density or collisionality so that RMPs are directly involved in ELM control.
- Refuelling of density pump out of the size of 30% requires an increase of the nominal fuelling rate by a factor of two.
- Immediately after the pellet plasma density decays on a fast time scale and during this phase the loss is dominated by ELMs, with a clear non diffusive character. The related normalised pellet retention time is similar to that in MAST.
- Pellets do not trigger bursts of ELMs.
- Energy confinement and pedestal temperatures are not restored to pre-RMP values by pellet refuelling.

These data provide a good starting point for future integrated density and ELM control experiments. We need to clarify why even at ITER-like collissionality, fuelling and amplitude of RMP field, the normalised ELM size is still larger than the ITER target. Also shallower pellet deposition should be tried to approach the ITER situation even closer. Inability of pellet refuelling to restore the energy confinement is another issue. It is possible that the situation discussed in this paper is the result of dominant ion heating and experiments to clarify this would be valuable.

**Acknowledgements**
This work has been carried out within the framework of the EUROfusion Consortium and has received funding from the Euratom research and training programme 2014-2018 under grant agreement No 633053 and from the RCUK Energy Programme [grant number EP/I501045]. To obtain further information on the data and models underlying this paper please contact PublicationsManager@ccfe.ac.uk. The views and opinions expressed herein do not necessarily reflect those of the European Commission. Authors would like to thank Drs C Maggi, H Meyer and anonymous referees for their valuable comments.